\pdfoutput=1
\documentclass[11pt]{article}

\usepackage{acl}

\usepackage{times}
\usepackage{latexsym}

\usepackage[T1]{fontenc}
\usepackage[utf8]{inputenc}
\usepackage{microtype}
\usepackage{inconsolata}
\usepackage{multirow}
\usepackage{booktabs}
\usepackage{geometry}
\usepackage{graphicx}
\usepackage{amsmath}
\usepackage{makecell}
\usepackage{amsfonts}
\usepackage{utfsym}
\usepackage{arydshln}
\usepackage[most]{tcolorbox}
\usepackage{CJKutf8}

\title{SLAM-Omni: Timbre-Controllable Voice Interaction System \\ with Single-Stage Training}

\author{
\textbf{Wenxi Chen\textsuperscript{\rm 1*}},
\textbf{Ziyang Ma\textsuperscript{\rm 1}},
\textbf{Ruiqi Yan\textsuperscript{\rm 1}},
\textbf{Yuzhe Liang\textsuperscript{\rm 1}},
\textbf{Xiquan Li\textsuperscript{\rm 1}}
\\
\textbf{Ruiyang Xu\textsuperscript{\rm 1}},
\textbf{Zhikang Niu\textsuperscript{\rm 1}},
\textbf{Yanqiao Zhu\textsuperscript{\rm 1}},
\textbf{Yifan Yang\textsuperscript{\rm 1}},
\textbf{Zhanxun Liu\textsuperscript{\rm 1}}
\\
\textbf{Kai Yu\textsuperscript{\rm 1}},
\textbf{Yuxuan Hu\textsuperscript{\rm 2}},
\textbf{Jinyu Li\textsuperscript{\rm 2}},
\textbf{Yan Lu\textsuperscript{\rm 2}},
\textbf{Shujie Liu\textsuperscript{\rm 2$^\dag$}},
\textbf{Xie Chen\textsuperscript{\rm 1$^\dag$}}
\\
 \textsuperscript{\rm1}MoE Key Lab of Artificial Intelligence, X-LANCE Lab, Shanghai Jiao Tong University \\
 \textsuperscript{\rm2}Microsoft Corporation
\\
 \texttt{\{1029713857,chenxie95\}@sjtu.edu.cn} \\
}

\usepackage{lipsum}
\newcommand\blfootnote[1]{%
  \begingroup
  \renewcommand\thefootnote{}\footnote{#1}%
  \addtocounter{footnote}{-1}%
  \endgroup
}

\begin{document}
\maketitle
\blfootnote{\hspace*{-0.3em}*This work was conducted during an internship at Microsoft Research Asia.}
\blfootnote{\hspace*{-0.3em}$^\dag$Corresponding authors.}

\begin{abstract}
Recent advancements highlight the potential of end-to-end real-time spoken dialogue systems, showcasing their low latency and high quality. 
In this paper, we introduce SLAM-Omni, a timbre-controllable, end-to-end voice interaction system with single-stage training. 
SLAM-Omni achieves zero-shot timbre control by modeling spoken language with semantic tokens and decoupling speaker information to a vocoder. 
By predicting grouped speech semantic tokens at each step, our method significantly reduces the sequence length of audio tokens, accelerating both training and inference.
Additionally, we propose historical text prompting to compress dialogue history, facilitating efficient multi-round interactions. 
Comprehensive evaluations reveal that SLAM-Omni outperforms prior models of similar scale, requiring only 15 hours of training on 4 GPUs with limited data.
Notably, it is the first spoken dialogue system to achieve competitive performance with a single-stage training approach, eliminating the need for pre-training on TTS or ASR tasks.
Further experiments validate its multilingual and multi-turn dialogue capabilities on larger datasets.\footnote{Demo at \url{https://SLAM-Omni.github.io}}
\end{abstract}

\section{Introduction}

With the advent of large language models (LLMs), recent developments \citep{achiam2023gpt,dubey2024llama,yang2024qwen2} have showcased their powerful capabilities in textual conversation.
In spoken dialogue systems, however, traditional methods rely on a cascaded pipeline involving automatic speech recognition (ASR) to transcribe user input, LLMs to generate textual responses, and text-to-speech (TTS) models to produce audio outputs.
This design faces two major issues: (1) significantly increased interaction latency, and (2) reliance on text-based interaction, which overlooks rich non-verbal information in speech dialogue, such as emotions and prosody.
The release of GPT-4o~\cite {openai2024gpt4o} has underscored the potential of real-time spoken dialogue systems in delivering seamless interaction. 
In response, several open-source frameworks, including Moshi \citep{defossez2024moshi}, Mini-Omni \citep{xie2024mini,xie2024mini2}, and LLaMA-Omni \citep{fang2024llama},  
have been developed for effective end-to-end voice-based interaction.

Existing spoken dialogue models (SDMs) primarily model speech with discretized audio tokens. 
Some approaches \citep{fang2024llama,wang2024freeze} rely on text embeddings to guide audio token generation, which limits their ability to generate critical audio paralinguistic attributes such as emotion and prosody.
Others \citep{zeng2024scaling,zhang2024omniflatten,nguyen2024spirit} adopt interleaved arrangements of audio and text tokens to restructure language modeling, while increasing training costs. 
A third category \citep{xie2024mini,xie2024mini2,mitsui2024pslm} employs a parallel speech-text generation method, which aligns closely with ours, balancing the delivery of intrinsic audio attributes and consuming of computational burden.

\begin{figure*}[t]
    \centering
    \includegraphics[width=\textwidth]{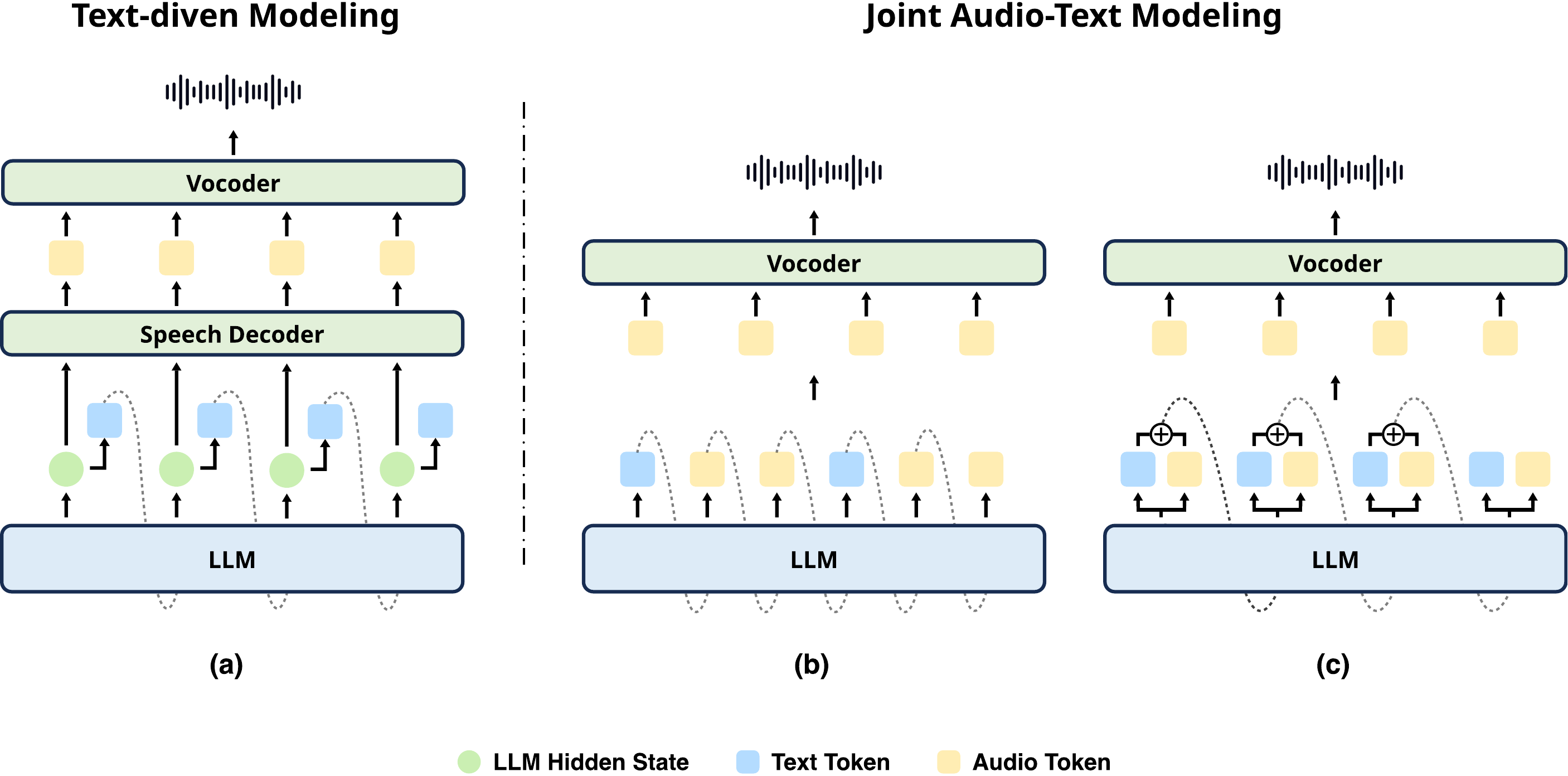}
    \caption{Illustration of existing end-to-end spoken dialogue modeling. (a): Text-driven modeling. (b): Interleaved audio-text modeling.  (c): Parallel audio-text modeling.}
    \label{fig:SDM-modeling}
\end{figure*}

A notable limitation of current SDMs is their disability to generate responses with diverse speaker timbres. 
This restriction primarily stems from the uniform timbre of responses in most training datasets and the lack of explicit speaker modeling in existing frameworks.
To address this gap, we propose the first zero-shot timbre control solution for dialogue systems. 
Drawing inspiration from zero-shot TTS \citep{wang2023neural}, our approach allows users to specify the desired output timbre by providing an audio prompt, paving the way for interactive applications such as personalized virtual assistants and customizable game character voices.

In this paper, we propose SLAM-Omni, a timbre-controllable, end-to-end spoken dialogue system with single-stage training.
For user speech input, the Whisper \citep{radford2023robust} encoder is employed to extract audio representations, which are then aligned with text embeddings via a projector and fed into the LLM.
On the output side, semantic audio tokens \citep{du2024cosyvoice} and text tokens are autoregressively predicted in parallel. These audio tokens naturally decouple speaker information into a separate vocoder, enabling zero-shot timbre control.
Inspired by VALL-E 2 \citep{chen2024vall}, SLAM-Omni predicts single-layer semantic tokens in grouped units per audio frame, reducing audio sequence length and accelerating training and inference. 
For multi-round spoken dialogue modeling, we introduce historical text prompting, which leverages text-only history rather than alternating audio-text streams. 
This strategy significantly compresses the dialogue history, improves data utilization, enables the model to handle more dialogue turns and enhances its instruction-following ability. 
During inference, instruction text is extracted from encoded audio embeddings with a Whisper decoder and response text is directly obtained from the generated text stream, both of which provide low-cost speech transcription that enables efficient multi-round voice interactions.
Comprehensive evaluations demonstrate that ASR or TTS pre-training is not necessary, while our SLAM-Omni, with only 15 hours of single-stage training on 4 GPUs, greatly outperforms prior models of similar scale in both speech content, quality and speech-text alignment.

Our contributions are summarized below:
\begin{itemize}
\item We propose the first zero-shot \textit{timbre control solution} for voice interaction systems with speaker-decoupled semantic tokens. 
\item \textit{Semantic Group Modeling} approach is proposed for accelerating single-layer semantic speech token generation and model training. 
\item \textit{Historical Text Prompting} is proposed for efficient multi-round history modeling in \hspace{-1mm} SDMs.
\item SLAM-Omni is the first voice assistant to achieve \textit{single-stage training}, requiring minimal data and computational resources.
\item  Experiments show that SLAM-Omni outperforms prior models of similar scale on text-related tasks, and shows superior performance on acoustic quality and speech-text alignment among all existing SDMs. 
Results on a larger dataset demonstrates its multilingual and multi-round dialogue capabilities. 
\end{itemize}

\begin{figure*}[t]
    \centering
    \includegraphics[width=\textwidth]{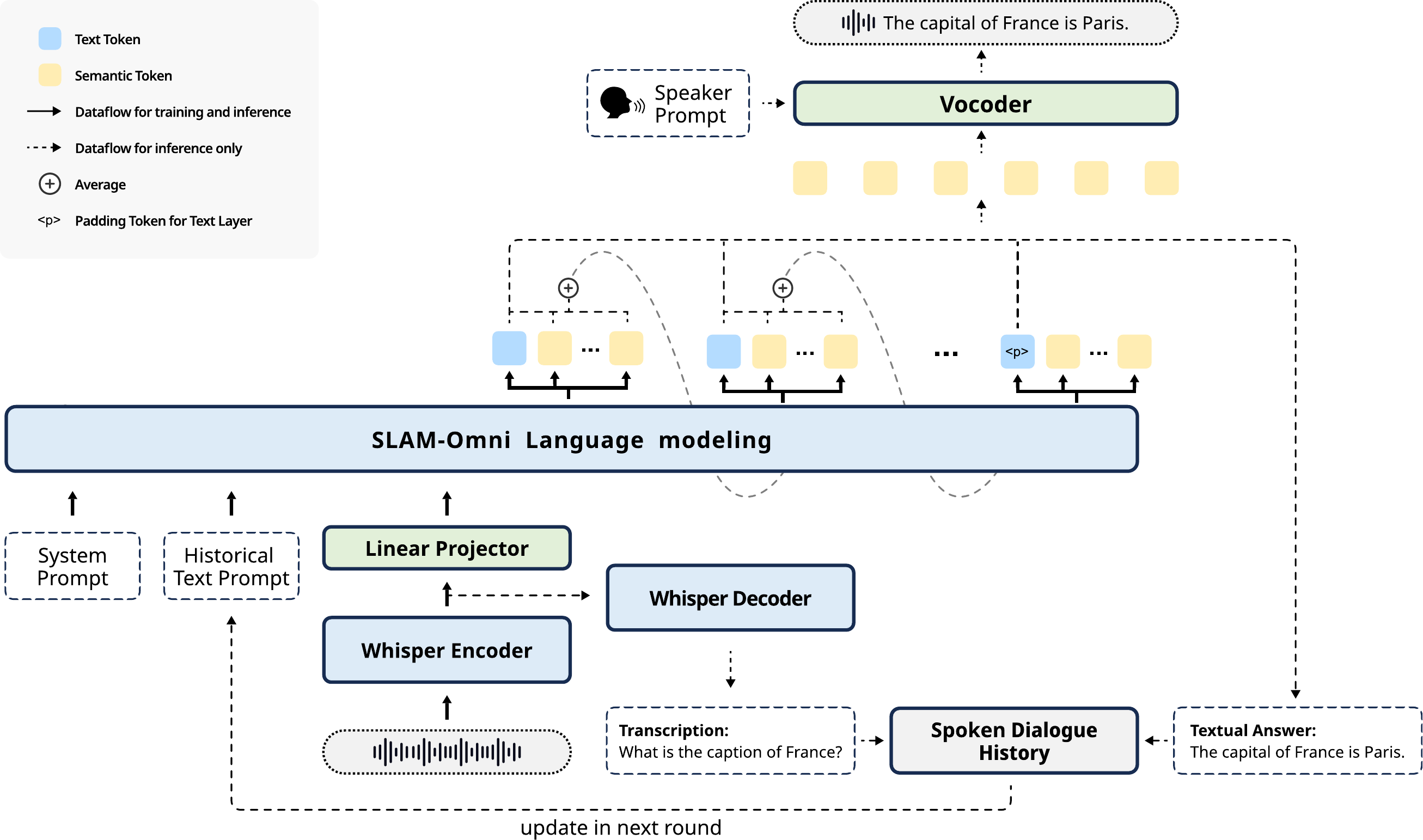}
    \caption{
    Overview of SLAM-Omni. System prompt, historical text prompt, followed by user speech embedding are concatenated as input for multi-turn voice interaction, while speaker prompt controls timbre using the vocoder. Semantic group modeling is used to accelerate speech token synthesis in the autoregressive language model.}
    \label{fig:SLAM-Omni}
\end{figure*}

\section{Related Work}

\subsection{End-to-End Spoken Dialogue Modeling}
\label{sec:modeling}

Existing end-to-end SDMs primarily model voice interaction by treating text as either an intermediate output or a hidden state to leverage the pre-trained knowledge of LLMs. 
As illustrated in Figure~\ref{fig:SDM-modeling}, these methods can be categorized into text-driven modeling and joint audio-text modeling.
For text-driven modeling, as shown in Figure~\ref{fig:SDM-modeling}a, existing methods \citep{fang2024llama,wang2024freeze} keep the original architecture of LLMs to retain textual abilities, using their hidden states as input to a speech decoder for audio generation. 
This approach effectively preserves LLMs knowledge but struggles to capture rich audio paralinguistic attributes such as emotion and prosody, since only text tokens are used for autoregressive modeling.
Joint audio-text modeling, illustrated in Figure~\ref{fig:SDM-modeling}b and c, is further divided into interleaved and parallel paradigms. Both paradigms incorporate audio tokens into the autoregressive modeling, theoretically enhancing the ability to model non-verbal information.
In the interleaved paradigm, models \citep{zhang2024omniflatten,zeng2024scaling,nguyen2024spirit} alternate between text and audio tokens during generation. 
This method typically requires extensive interleaved speech-text data and pre-training for re-modeling LLMs. 
In contrast, the parallel paradigm, adopted by models like PSLM \citep{mitsui2024pslm}, Mini-Omni \citep{xie2024mini,xie2024mini2}, and our proposed SLAM-Omni, employs autoregressive modeling of text and audio tokens in parallel. 
However, unlike PSLM and Mini-Omni, SLAM-Omni predicts single-layer grouped semantic tokens to accelerate audio generation process. 
Combining semantic group modeling with single-stage training, we achieve an end-to-end SDM built on a pre-trained LLM that requires significantly less training costs compared to previous solutions.

\subsection{Speech Tokenization}
Speech tokenization is a foundational technique in speech language models (SLMs), typically categorized into acoustic tokens and semantic tokens \citep{zhang2023speechgpt,borsos2023audiolm}.
Acoustic tokens, derived from neural audio codecs \citep{defossez2022high,zeghidour2021soundstream} and optimized for reconstructing high-quality audio, have been widely adopted in SLMs for speech synthesis and editing \citep{wang2023neural,peng2024voicecraft}, as well as in SDMs for voice interaction \citep{xie2024mini,xie2024mini2,wang2024freeze}.
In contrast, semantic tokens are obtained by discretizing speech representations extracted from self-supervised speech pre-trained models \citep{hsu2021hubert,chung2021w2v}, focusing on capturing semantic content rather than acoustic detail.
These tokens are also extensively used in SLMs \citep{an2024funaudiollm,ma2024language} and SDMs \citep{zeng2024glm,fang2024llama}. 
Among these approaches, CosyVoice \citep{du2024cosyvoice} leverages supervised semantic tokens to enable zero-shot TTS, demonstrating the potential of semantic tokens for timbre control. This insight inspires our work, which seeks to extend such functionality to SDMs—a promising yet underexplored direction in the field.

\section{SLAM-Omni}
\subsection{Overview}
As shown in Figure~\ref{fig:SLAM-Omni}, SLAM-Omni processes input speech using continuous features and adopts parallel audio-text modeling with discrete semantic audio tokens for speech output. 
This section details its modeling strategies, covering speech input, speech output, timbre control, and multi-round spoken dialogue, along with its training methodology.

\subsection{Speech Input Modeling}
SLAM-Omni employs the Whisper encoder \citep{radford2023robust} to extract audio features $\mathbf{A} = [a_1, a_2, \cdots, a_N]$ from user speech instructions at a frequency of 50 Hz.
Whisper, a speech recognition model trained on large-scale supervised cross-lingual speech data, provides precise transcription and robust multilingual support, serving as a foundational component for SLAM-Omni's multi-turn and multilingual dialogue capabilities.
Following \citet{ma2024embarrassingly}, we downsample $\mathbf{A}$ by concatenating every $k$ consecutive frames along the feature dimension, yielding intermediate features $\mathbf{A}^I = [a^I_1, a^I_2, \dots, a^I_{N'}]$, where $a^I_i = a_{(i-1)*k+1} \oplus a_{(i-1)*k+2} \oplus \cdots \oplus a_{i*k-1}$ and $N' = N // k$.
A linear encoder projector then transforms $\mathbf{A}^I$ into $\mathbf{A}^P$ to ensure alignment with LLM’s embedding dimension, defined as $\mathbf{A^P} = \text{MLP}(\mathbf{A}^I)$. These reduced speech features are concatenated with the prompt embeddings $\mathbf{P}$ and serve as input to the LLM.

\subsection{Semantic Group Modeling}
\label{sec:semantic-group}

For speech output, we adopt parallel audio-text modeling, predicting single-layer semantic tokens \citep{du2024cosyvoice} alongside text tokens autoregressively. 
To achieve this, the original LLM vocabulary \( V_t \) and embedding space are extended with a new codebook \( V_a \) for audio tokens, resulting in an expanded vocabulary \( V_j = V_t \cup V_a \). The original word embedding matrix is preserved, while the embeddings for audio tokens are randomly initialized.

\begin{figure}[htbp]
    \centering
    \includegraphics[width=0.48\textwidth]{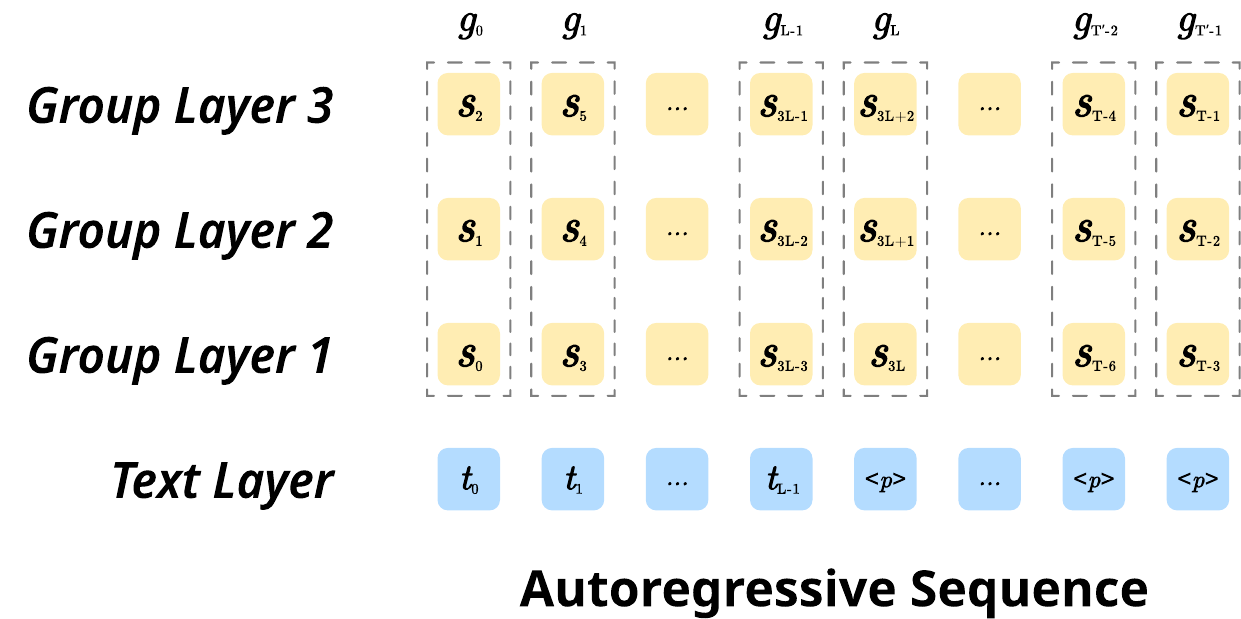}
    \caption{Illustration of \textit{semantic group modeling} with \( G = 3\). At each step of the autoregressive process, embeddings of grouped semantic tokens and text tokens are aggregated as the input to the LLMs.}
    \label{fig:group-prediction}
\end{figure}

At each generation step, the LLM outputs logits \( L_j \in \mathbb{R}^{|V_j|} \), which are partitioned into \( L_t \in \mathbb{R}^{|V_t|} \) and \( L_a \in \mathbb{R}^{|V_a|} \), representing predicted distributions for text and audio tokens, respectively. However, generating text and audio tokens at the same rate introduces a key challenge: 
there is a substantial frequency mismatch between text tokens (\textasciitilde3Hz) and semantic tokens (50Hz). 
The high frequency of audio tokens results in considerably longer sequences, significantly increasing both training and inference costs, as well as leading to higher latency in real-time speech generation.

To mitigate these issues, we propose \textit{semantic group modeling}, which allows the model to predict multiple audio tokens simultaneously at each step, as illustrated in Figure \ref{fig:group-prediction}.
This approach projects the audio logits \( L_a \) into group-sized logits \( L_g \) with a linear layer, where \( L_g \in \mathbb{R}^{|V_a| \times G} \), and \( G \) denotes the group size. 
During training, the original semantic token sequence \( \mathbf{S}^T = [s_0, s_1, \dots, s_{T-1}] \) is grouped as \( \mathbf{G}^T = [g_0, g_1, \dots, g_{T'-1}] \), where:

{\small
\[
g_i = [s_{i \cdot G}, s_{i \cdot G + 1}, \dots, s_{(i+1) \cdot G - 1}], \quad T' = T // G . \tag{1}
\]
}

Given prompt embeddings \( \mathbf{P} \), audio features \( \mathbf{A}^P \) 
and text token sequence \( \mathbf{T}^L = [t_0, t_1, \dots, t_{L-1}] \), the training objective is defined as a weighted cross-entropy loss:

{\small
\[
\mathcal{L} = \lambda_{\text{text}} \mathcal{L}_{\text{text}} + \lambda_{\text{audio}} \mathcal{L}_{\text{audio}} \tag{2}
\]
}

where:
{\small
\[
\mathcal{L}_{\text{text}} = -\frac{1}{L} \sum_{i=1}^{L} \log p(t_i \mid \mathbf{P}, \mathbf{A}^P, \mathbf{G}^T_{<i}, \mathbf{T}^L_{<i})  \tag{3}
\]
}
{\small
\[
\mathcal{L}_{\text{audio}} = -\frac{1}{T'G}  \sum_{i=1}^{T'} \sum_{j=1}^{G} \log p(s_{i \cdot G + j} \mid \mathbf{P}, \mathbf{A}^P, \mathbf{G}^T_{<i}, \mathbf{T}^L_{<i})  \tag{4}
\]
}

Here, \( \mathcal{L}_{\text{text}} \) and \( \mathcal{L}_{\text{audio}} \) represent the losses for text and audio token predictions, respectively, while \( \lambda_{\text{text}} \) and \( \lambda_{\text{audio}} \) are 
corresponding weights.

\subsection{Controllable Timbre Modeling}
Previous approaches disentangle speech by modeling distinct subspaces for different attributes \citep{ju2024naturalspeech} or predicting supervised semantic tokens that separate content and speaker information \citep{du2024cosyvoice}. 
These methods enable timbre disentanglement from semantic content, achieving zero-shot TTS where users can freely adjust the system’s vocal timbre by providing audio prompts.

Building on these insights from TTS modeling, we extend zero-shot timbre control to SDMs. 
By modeling speech content as semantic tokens, SLAM-Omni inherently disentangles timbre from linguistic information.
Following techniques demonstrated in zero-shot TTS (e.g., CosyVoice), we employ a conditional flow matching model to convert semantic tokens and speaker prompts into mel spectrograms, which are then synthesized into waveforms via HiFi-GAN \citep{kong2020hifi}.
For real-time speech generation, same as common practice like~\citet{zeng2024scaling}, block causal attention is adopted in the Transformer of flow matching.

\subsection{Historical Text Prompting}
Previous multi-turn spoken dialogue modeling often interleave text and audio tokens as the LLM history \citep{wang2024freeze,zeng2024glm}. 
However, the lengthy audio token sequences pose challenges for model training, especially in joint audio-text modeling requiring full fine-tuning, significantly increasing computational costs and limiting the number of dialogue turns. 
Moreover, longer histories hinder in-context learning and raise the risk of forgetting earlier dialogue content.

To address these issues, we introduce \textit{Historical Text Prompting}, which exclusively utilizes text modality to represent dialogue history. 
As shown in Figure~\ref{fig:SLAM-Omni}, SLAM-Omni structures multi-turn interactions using the template: <System> <History> <Input> <Answer>.
Here, the system prompt specifies the model's role and the dialogue task, while the history prompt stores past dialogue content in text form.
This approach aligns naturally with the training paradigm of LLMs, inheriting their robust text-based in-context learning capabilities.
Moreover, it eliminates the burden of modeling long audio sequences as history, enabling the model to handle more dialogue turns within a constrained context window.

\begin{figure}[htbp]
\centering
\includegraphics[width=0.48\textwidth]{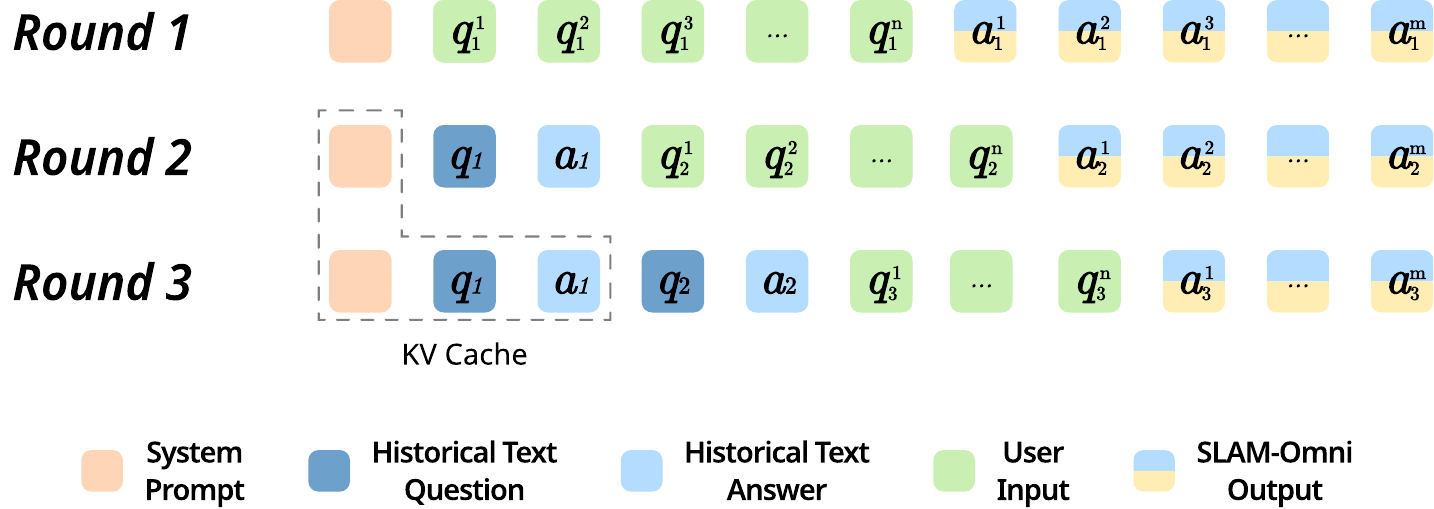}
\caption{Illustration of the key-value cache mechanism in \textit{Historical Text Prompting} for multi-round dialogue.
}
\label{fig:kv-cache}
\end{figure}

During inference, speech features \( \mathbf{A} \) extracted by Whisper can be decoded into the transcription of the input speech, represented as \(\text{Decoder}(\mathbf{A})\). 
On the output side, the generated text tokens are converted back into text using the tokenizer.
Both the textual question and answer are appended to the dialogue history for subsequent turns.
As illustrated in Figure \ref{fig:kv-cache}, the transcription of the first-round spoken dialogue is incorporated into the historical prompt. 
During the second round of inference, the corresponding key-value cache is generated and can be reused in the third and subsequent rounds of dialogue, facilitating efficient multi-round inference.

\subsection{Single-Stage Training}
Current spoken dialogue models typically depend on multi-stage training, including modality adaptation, modality alignment, and supervised fine-tuning \citep{ji2024wavchat}. 
These designs demand intricate training strategies, such as coordinating module training across stages and tuning numerous hyperparameters, leading to substantial time and computational overhead.

Aligned with the goal of making SDMs training accessible to everyone, SLAM-Omni achieves outstanding performance through one-stage training with minimal data. 
In our experiments, both TTS and ASR training exhibit rapid loss convergence (see Appendix \ref{app:pre-train_details}), underscoring that extensive modality alignment pre-training is unnecessary in our modeling method.
Moreover, further experiments reveal that pre-training negatively impacts model's ability to follow instructions and retain general knowledge, as detailed in Section \ref{sec:training_strategy}.

\section{Experimental Setup}
\subsection{Datasets}
\label{sec:datasets}

\begin{table}[htbp]
\small
\centering
\resizebox{1\linewidth}{!}{
\begin{tabular}{ccccc}
\toprule
\textbf{\makecell{Data Source}} & 
\textbf{\makecell{Multi-turn}} & \textbf{\makecell{Instruction \\ Duration}} & \textbf{\makecell{Response \\ Duration}} & \textbf{\makecell{\#Samples}} \\
\midrule
VoiceAssistant-400K & \usym{2717} & 664 h & 3,234 h & 460K \\ 
UltraChat & \usym{2713} & 619 h & 1,951 h & 300K \\
Belle\_train\_3.5M\_CN & \usym{2713} & 2,488 h & 6,418 h & 1.4M \\
\bottomrule
\end{tabular}
}
\caption{The statistics of training datasets.}
\label{tab:training-dataset}
\end{table}

As most publicly available dialogue datasets are text-based, we synthesize spoken dialogue corpora using zero-shot TTS systems. 
Specifically, we utilize discrete speech tokens from \citet{du2024cosyvoice} and employ CosyVoice\footnote{\scriptsize\url{https://github.com/FunAudioLLM/CosyVoice}} to generate dialogue utterances. 
For user inputs, the CosyVoice-300M model is employed to produce corresponding speech.
Vocal timbre is controlled by randomly sampling speaker prompts from a timbre library, which contains 1007 English and 1010 Chinese human audio prompts sourced from seed-tts-eval\footnote{\scriptsize\url{https://github.com/BytedanceSpeech/seed-tts-eval}} \citep{anastassiou2024seed}.
For assistant responses, we use the text-to-token LLM from CosyVoice-300M-SFT to generate semantic tokens, which are used as target audio tokens during SLAM-Omni training.

Table~\ref{tab:training-dataset} summarizes the datasets used to synthesize spoken dialogue corpora.
The training data include VoiceAssistant-400K\footnote{\scriptsize\url{https://huggingface.co/datasets/gpt-omni/VoiceAssistant-400K}} from Mini-Omni \citep{xie2024mini}, the English multi-turn dataset UltraChat\footnote{\scriptsize\url{https://huggingface.co/datasets/stingning/ultrachat}} \citep{ding2023enhancing}, and the Chinese dialogue dataset Belle\_train\_3.5M\_CN\footnote{\scriptsize\url{https://huggingface.co/datasets/BelleGroup/train_3.5M_CN}} \citep{ji2023exploring}. 
We clean the synthesized data by removing written artifacts (e.g., emojis, URLs), and we limit the duration of instructions and responses to a maximum of 30 and 60 seconds, respectively, to better align with natural conversational scenarios.
For the primary experiments with SLAM-Omni, only VoiceAssistant-400K is used, while the remaining datasets are incorporated in supplementary experiments to evaluate the model’s performance in multi-turn and multilingual dialogue tasks.

\subsection{Training and Inference Details}
\label{sec:config}

To ensure a fair comparison in low-resource settings, particularly with Mini-Omni \citep{xie2024mini,xie2024mini2}, another parallel audio-text modeling approach, we utilize Qwen2-0.5B\footnote{\scriptsize\url{https://huggingface.co/Qwen/Qwen2-0.5B}}
\citep{yang2024qwen2} as the LLM backbone and Whisper-small\footnote{\scriptsize\url{https://huggingface.co/openai/whisper-small}} \citep{radford2023robust} as the speech encoder and decoder.
Following \citet{ma2024embarrassingly}, user speech instructions are zero-padded to 30 seconds before being processed by the Whisper encoder, with the resulting speech features downsampled using \( k = 5 \).
In the main experiments, SLAM-Omni adopts a semantic group size of \( G = 3 \). 
For ablation studies on group size, models with \( G > 1 \) include an additional linear layer for predicting grouped tokens.

During single-stage training, SLAM-Omni undergoes full fine-tuning, with the Whisper encoder kept frozen.
The weights for \( \mathcal{L}_{\text{text}} \) and \( \mathcal{L}_{\text{audio}} \) are set to 1. 
We use the AdamW optimizer \citep{loshchilov2017decoupled} with a peak learning rate of \( 1 \times 10^{-4} \) and a batch size of 24. 
Training spans 100,000 steps, with the first 1,000 steps used for warmup, followed by a linear decay schedule.
A validation set comprising 1\% of the training data is used, and validation is performed every 3,000 updates, saving checkpoints based on the lowest validation loss. 
For a direct comparison with Mini-Omni, our primary experiments are \textbf{only conducted on VoiceAssistant-400K}, a subset of Mini-Omni's training data. Details on multilingual and multi-turn training are provided in Appendix \ref{app:multi-round} and Appendix \ref{app:chinese}. 
The entire training process takes approximately 15 hours on 4 NVIDIA A100 GPUs.

For inference, we use greedy search decoding with a repetition penalty of 1.2 applied to both audio and text layers. 
Consistent with \citep{fang2024llama}, models are evaluated using non-streaming decoding for speech response generation.

\begin{table}[htbp]
\small
\centering
\resizebox{1\linewidth}{!}{
\begin{tabular}{ccccc}
\toprule
\textbf{Types} & \textbf{Datasets} & \textbf{\#Samples} & \textbf{Avg. \#Words} & \textbf{Avg. Audio len} \\
\midrule
\multirow{2}{*}{Understanding} 
& Repeat & 252 & 21.76 & 8.04  \\ 
& Summary & 118 & 58.93 & 20.38  \\
\midrule
\multirow{3}{*}{Reasoning} 
& StoralEval & 201 & 66.46 & 20.52  \\ 
& TruthfulEval & 470 & 10.87 & 3.40  \\ 
& MLC & 177 & 22.43 & 7.56 \\
\midrule
\multirow{3}{*}{\makecell{Oral \\ Conversation}} 
& AlpacaEval & 199 & 16.37 & 5.67  \\ 
& CommonEval & 200 & 8.16 & 4.83  \\ 
& WildchatEval & 349 & 14.68 & 4.75  \\
\bottomrule
\end{tabular}
}
\caption{The statistics of main evaluation datasets.}
\label{tab:evaluation-dataset}
\end{table}

\begin{table*}[htbp]
\scriptsize
\centering
\resizebox{1\linewidth}{!}{
\begin{tabular}{lcccccccccc}
\toprule
\multirow{3}{*}{\textbf{Models}} & \multirow{3}{*}{\makecell{\textbf{LLM} \\ \textbf{Scale}}} 
 & \multicolumn{2}{c}{\textbf{Understanding}} & \multicolumn{3}{c}{\textbf{Reasoning}} & \multicolumn{3}{c}{\textbf{Oral Conversation}} & \multirow{3}{*}{\textbf{Overall}} \\
\cmidrule(lr){3-4}
\cmidrule(lr){5-7}
\cmidrule(lr){8-10}
 &  & Repeat & Summary & StoralEval & TruthfulEval & MLC & AlpacaEval & CommonEval & WildchatEval \\
\midrule
\textcolor{gray}{Qwen2-7B-instruct$^\dag$} &  \textcolor{gray}{7B} & 
\textcolor{gray}{96.87} & \textcolor{gray}{97.45} & \textcolor{gray}{82.35} & \textcolor{gray}{67.89} & \textcolor{gray}{73.26} & \textcolor{gray}{95.91} & \textcolor{gray}{85.93} & \textcolor{gray}{92.72} & \textcolor{gray}{86.55} \\
\textcolor{gray}{Freeze-Omni} & \textcolor{gray}{7B} &  \textcolor{gray}{70.89} & \textcolor{gray}{78.87} & \textcolor{gray}{57.74} & \textcolor{gray}{46.95} & \textcolor{gray}{42.56} & \textcolor{gray}{52.23} & \textcolor{gray}{48.70} & \textcolor{gray}{55.80} & \textcolor{gray}{56.72} \\
\textcolor{gray}{LLaMA-Omni} & \textcolor{gray}{8B} &  \textcolor{gray}{45.62} & \textcolor{gray}{80.68} & \textcolor{gray}{50.65} & \textcolor{gray}{45.13} & \textcolor{gray}{44.44} & \textcolor{gray}{64.36} & \textcolor{gray}{58.40} & \textcolor{gray}{72.19} & \textcolor{gray}{57.68} \\
\textcolor{gray}{GLM-4-Voice} & \textcolor{gray}{9B} & 
\textcolor{gray}{90.95} & \textcolor{gray}{91.07} & \textcolor{gray}{73.80} & \textcolor{gray}{59.28} & \textcolor{gray}{57.82} & \textcolor{gray}{80.77} & \textcolor{gray}{63.07} & \textcolor{gray}{78.76} & \textcolor{gray}{74.44} \\
\hdashline
\textcolor{gray}{Qwen2-0.5B-instruct$^\dag$} &  \textcolor{gray}{0.5B} & 
\textcolor{gray}{60.12} & \textcolor{gray}{78.59} & \textcolor{gray}{49.82} & \textcolor{gray}{39.73} & \textcolor{gray}{52.92} & \textcolor{gray}{58.93} & \textcolor{gray}{57.50} & \textcolor{gray}{63.97} & \textcolor{gray}{57.70} \\
Mini-Omni & 0.5B & 5.07 & 32.20 & 23.25 & 25.06 & 2.82 & 30.99 & 29.80 & 31.42 & 22.58 \\
Mini-Omni2 & 0.5B & 8.10 & 40.06 & 28.49 & 26.92 & 6.97 & 34.81 & 30.70 & 36.43 & 26.56 \\
\textbf{SLAM-Omni (ours)} & 0.5B & 
\textbf{12.26} & \textbf{66.21} & \textbf{36.95} & \textbf{34.65} & \textbf{21.85} & \textbf{48.98} & \textbf{41.03} & \textbf{52.61} & \textbf{39.32} \\
\bottomrule
\end{tabular}
}
\caption{ChatGPT scores of SDMs and LLMs across three dimensions.
$^\dag$The Qwen2 series models are text-based, single-modal LLMs, with transcription input generated by Whisper-large-v3.
}
\label{tab:main-chatgpt}
\end{table*}

\subsection{Evaluation for Spoken Dialogue Models}
\label{sec:Evaluation for Spoken Dialogue Models}
Previous SDMs lacked a thorough evaluation of voice interaction capabilities. 
VoiceBench \citep{chen2024voicebench} is the first benchmark for voice assistants, but it only assesses the model's text output.
To bridge this gap, we propose a comprehensive evaluation framework that directly measures the speech-to-speech capabilities of SDMs. Voice interaction in SDMs can be broken down into three key stages: understanding, reasoning, and oral conversation. 
We have designed eight distinct test sets that assess SDMs across these three dimensions:

\paragraph{Understanding} To evaluate the model's ability of comprehending and following user instructions, we build two datasets to require the model to repeat the user's words or summarize a story.

\paragraph{Reasoning}
We adapt samples from TruthfulQA \citep{lin2021truthfulqa} and STORAL \citep{guan2022corpus}, and design additional questions on math, logic, and common sense (MLC) to assess the model's general knowledge and reasoning ability. 

\paragraph{Oral Conversation }
We use AlpacaEval \citep{li2023alpacaeval} and CommonEval \citep{ardila2019common} from VoiceBench, along with real-life questions from WildChat \citep{zhao2024wildchat}, to test the model's conversational ability in open-ended scenarios.

The model's inference results on these tasks are evaluated using the following metrics:

\paragraph{ChatGPT Score} 
To assess the \textbf{content quality} of the model's responses, we use Whisper-large-v3\footnote{\scriptsize\url{https://huggingface.co/openai/whisper-large-v3}} to transcribe the speech output into text, followed by evaluation using GPT-4o mini \citep{openai2024gpt4omini}. 
The model is prompted to score the transcription based on predefined criteria, including accuracy, relevance, clarity, and completeness, with detailed prompts provided in Appendix \ref{app:scoring-criteria}.

\paragraph{UTMOS Score} 
To measure the overall \textbf{speech quality}, we use the UTMOS \citep{saeki2022utmos} model to predict mean opinion scores (MOS).

\paragraph{WER Score} 
To evaluate the \textbf{speech-text alignment}, we calculate the word error rate (WER) between the speech transcription and the corresponding text response, referred to as ASR-WER.

The overall scores for UTMOS and ASR-WER are calculated as the average of their respective scores across these eight evaluation datasets.

Table~\ref{tab:evaluation-dataset} summarizes the evaluation datasets, with details and scoring criteria in Appendices \ref{app:evaluation} and \ref{app:scoring-criteria}. Descriptions for the multi-turn and Chinese evaluation datasets are in Appendices \ref{app:multi-round} and \ref{app:chinese}.
We assess SLAM-Omni alongside the Mini-Omni  \citep{xie2024mini,xie2024mini2}, both using a 0.5B LLM backbone, and compare against larger SDMs including Freeze-Omni \citep{wang2024freeze}, Llama-Omni \citep{fang2024llama}, and GLM-4-Voice \citep{zeng2024glm}, as well as LLMs such as Qwen2-0.5B-instruct and Qwen2-7B-instruct \citep{yang2024qwen2}.

\begin{table}[htbp]
\centering
\small
\resizebox{1\linewidth}{!}{
\begin{tabular}{lccc}
\toprule
\textbf{Models} & \textbf{ChatGPT Score $\uparrow$} & \textbf{UTMOS $\uparrow$} & \textbf{ASR-WER $\downarrow$}  \\
\midrule
\textcolor{gray}{Freeze-Omni} & \textcolor{gray}{56.72} & \textcolor{gray}{4.37} & \textcolor{gray}{16.32\%} \\
\textcolor{gray}{LLaMA-Omni} & \textcolor{gray}{57.68} & \textcolor{gray}{4.02} & \textcolor{gray}{10.42\%} \\
\textcolor{gray}{GLM-4-Voice} & \textcolor{gray}{74.44} & \textcolor{gray}{4.15} & \textcolor{gray}{12.71\%} \\
\hdashline
Mini-Omni & 22.58 & 4.42 & 6.05\% \\
Mini-Omni2 & 26.56 & 4.43 & 10.24\% \\
\textbf{SLAM-Omni (ours)} & \textbf{39.32} & \textbf{4.45} & \textbf{4.54\%} \\
\bottomrule
\end{tabular}
}
\caption{Overall evaluation results for SDMs.}
\label{tab:main-overall}
\end{table}

\section{Experimental Results}

\subsection{Main Results}
Tables \ref{tab:main-chatgpt} and \ref{tab:main-overall} present the performance of SLAM-Omni compared to mainstream SDMs.
Given our focus on low-resource settings, we mainly benchmark performance against models with the same size, while including larger-scale SDMs and LLMs in \textcolor{gray}{gray} as references.
Results show that, despite SLAM-Omni’s single-stage training on only the third-phase Mini-Omni data, it significantly improves speech content, audio quality, and speech-text alignment.
Although gaps in textural abilities exist compared to larger SDMs (which we believe derives from the pre-trained LLM model size), SLAM-Omni notably surpasses them in UTMOS and ASR-WER scores, demonstrating its advantages in audio modeling.
Further assessments of multi-turn spoken dialogues and performance on Chinese voice interactions are detailed in Appendices \ref{app:multi-round} and \ref{app:chinese}, respectively.

In ChatGPT-based evaluations, SLAM-Omni surpasses Mini-Omni in understanding, reasoning, and oral conversation, indicating that it preserves more pre-trained LLM knowledge and instruction-following capabilities.
However, it still falls short of Qwen2-0.5B-instruct. Although both models are fine-tuned from Qwen2-0.5B-base, Qwen2-0.5B-instruct benefits from extensive text-based instruction tuning, whereas SLAM-Omni relies solely on a 400K spoken-dialogue dataset.
Evaluations of larger-scale models reveal that current SDMs consistently underperform relative to similarly sized LLMs.
One possible reason for this disparity is the relatively limited exploration of data during SDMs training compared to the extensive pre-training, SFT, and RLHF undertaken for LLMs. 
How to effectively preserve, or even enhance, the original knowledge of the LLM while incorporating spoken dialogue data during SDMs training remains a promising and important research direction.

In terms of audio quality and speech-text alignment, SLAM-Omni surpasses all other SDMs, particularly on ASR-WER metrics, which may be attributed to our semantic group modeling strategy. 
By leveraging grouped semantic tokens, SLAM-Omni achieves tighter speech-text alignment, ensuring that the generated audio closely matches its textual counterpart.
In contrast, larger SDMs often generate audio that fails to align with their intermediate textual outputs, as evidenced by their ASR-WER exceeding 10\%.
More specifically, these models struggles with long-form content generation, with sometimes audio generation interrupted midway, or extended silence generated. 
These issues ultimately lower their UTMOS and ASR-WER scores in our evaluations.

\subsection{Multi-turn Interaction}
Appendix~\ref{app:multi-round} details the multi-turn spoken dialogues settings and results. 
Our experiments suggests that exposing the model to multi-turn spoken dialogues with historical text prompting can activate its underlying textual in-context learning capabilities. 
As a result, even though the model was fine-tuned exclusively on spoken instructions, it can effectively interpret textual instructions.

\subsection{Ablation Study}
We conduct ablation studies to further validate the efficiency and  effectiveness of our modeling and training strategy.
All experiments were conducted on 4 NVIDIA A100 GPUs for fair comparisons.

\begin{table}[htbp]
\centering
\small
\resizebox{1\linewidth}{!}{
\begin{tabular}{ccccc}
\toprule
\textbf{Group Size \( G \)} & \makecell{\textbf{ChatGPT} \\ \textbf{Score}}$\uparrow $ & \textbf{UTMOS} $\uparrow$ & \textbf{ASR-WER} $\downarrow$ & \makecell{\textbf{GPU} \\ \textbf{Hours}}   \\
\midrule
1  & 34.17 & 4.44 & 18.23\% & 126   \\
2  & 35.22 & \textbf{4.46} & 8.00\%  & 78    \\
3  & \textbf{39.32} & 4.45 & 4.54\%  & 60    \\
4  & 37.19 & 4.45 & \textbf{4.31\%}  & 52    \\
5  & 33.93 & 4.43 & 4.85\% & 50  \\
\bottomrule
\end{tabular}
}
\caption{Ablation study for the group size \( G \).}
\label{tab:ablation-group_size}
\end{table}

\subsubsection{Effect of Group Size}
Table \ref{tab:ablation-group_size} presents the impact of different group sizes in semantic group modeling on model performance. 
The results indicate that semantic group modeling significantly enhances the model's speech-text alignment and enables it to generate more helpful responses. 
Specifically, when \( G \geq 3 \), the model achieves an ASR-WER below 5\%, whereas the model without grouping semantic tokens (\( G = 1 \)) shows a much higher ASR-WER of 18.23\%. 
This gap arises primarily due to the frequency mismatch between audio tokens and text tokens, as discussed in Section \ref{sec:semantic-group}. 
By properly reducing the length of audio sequences, semantic group modeling effectively alleviates this mismatch, enables better semantic alignment between audio and text tokens.
Moreover, it ensures better retention of pre-trained LLM knowledge after dialogue data fine-tuning, as evidenced by the improved ChatGPT scores.

Additionally, semantic group modeling substantially reduces training and inference costs. 
During training, a lightweight group prediction layer is employed to compresses audio sequences, drastically lowering GPU memory consumption and training overhead. 
As a result, the model achieves superior performance with less than half the GPU hours required by baselines. 
This approach also accelerates inference.
For instance, when using a streaming vocoder with chunk sizes of 30 tokens, a model with \( G = 3 \) requires only 10 LLM inference steps to produce the first audio packet.
This reduced latency ensures seamless audio generation, enhancing user experience in voice interactions.

\begin{table}[htbp]
\centering
\small
\resizebox{1\linewidth}{!}{
\begin{tabular}{lcccc}
\toprule
\textbf{Setting} & \makecell{\textbf{ChatGPT} \\ \textbf{Score} }$\uparrow$ & \textbf{UTMOS} $\uparrow$ & \textbf{ASR-WER} $\downarrow$  & \makecell{\textbf{GPU} \\ \textbf{Hours}} \\
\midrule
SLAM-Omni &  \textbf{39.32} & 4.45 & 4.54\% & 60 \\
- w/ ASR pre-training & 34.02 & 4.45 & \textbf{4.38}\% & 132 \\
- w/ TTS pre-training & 27.22 & \textbf{4.46} & 4.53\% & 160 \\
\bottomrule
\end{tabular}
}
\caption{Ablation study for training strategy.}
\label{tab:ablation-training_strategy}
\end{table}

\subsubsection{Training Strategy}
\label{sec:training_strategy}
Previous voice interaction systems typically rely on a multi-stage training pipeline, beginning with modality alignment pre-training tasks (e.g., ASR or TTS) before transitioning to fine-tuning on dialogue data.
However, as shown in Table~\ref{tab:ablation-training_strategy}, while ASR and TTS pre-training slightly improve audio-text alignment—evidenced by lower ASR-WER—they fail to enhance overall performance on spoken interactive tasks. 
In contrast, SLAM-Omni, trained using a single-stage strategy, significantly outperforms pre-trained models in ChatGPT scores while maintaining comparable audio quality.
One possible explanation is that focusing solely on a single pre-training task can diminish the model’s instruction-following capability and erode its general knowledge base.
In contrast, our experiments demonstrate that applying single-stage fine-tuning directly on speech-to-speech datasets helps SLAM-Omni retain more of the original LLM’s pre-trained knowledge. 
This streamlined approach also eliminates the need for a separate pre-training step and more than doubles the training efficiency.

\section{Conclusion}
In this work, we propose SLAM-Omni, a timbre-controllable, end-to-end spoken dialogue model with single-stage training. 
Through a novel semantic group modeling, SLAM-Omni effectively aligns audio and text modalities during audio generation, as well as accelerating both training and inference. 
Employing supervised semantic tokens to disentangle speaker information, SLAM-Omni is capable of zero-shot timbre control. 
To address the issues posed by long audio histories, we introduce historical text prompting technique, which stores dialogue history as text and uses key-value caches for efficient multi-turn inference. 
Despite limited data and only 60 GPU hours of training, SLAM-Omni surpasses previous SDMs of similar scale on text-related abilities, and exceeds all SDMs on acoustic quality and speech-text alignment.


\section*{Limitations}
There are two limitations to this work.
First, while historical text prompting effectively mitigates the burden of handling long audio sequences during training and inference, it sacrifices the rich non-verbal information accumulated from previous dialogue turns. 
In certain scenarios, retaining this historical context is crucial for maintaining dialogue coherence and depth. 
Further exploration is needed to efficiently retain such information in SDMs.
Second, although SLAM-Omni demonstrates efficient modeling for smaller-scale LLMs, extending this approach to larger LLMs remain to be explored. 
Unlike purely text-driven methods, joint audio-text modeling necessitates substantially more training data for large-scale models. 
Striking a balance between efficient audio-text joint modeling and minimizing the loss of the original LLM’s inherent knowledge remains a critical direction for future research.


\bibliography{custom}

\appendix

\newpage

\section{Pre-training Details}
\label{app:pre-train_details}
For ASR and TTS pre-training, we exclusively utilize the VoiceAssistant-400K dataset to ensure consistency and avoid introducing external data. 
During ASR pre-training, the speech instructions are provided as input, with their corresponding transcriptions serving as the target outputs. 
Conversely, for TTS pre-training, the transcriptions of the speech responses are used as input text, while the corresponding semantic tokens are set as the prediction targets.
The optimization and learning strategies align with those employed during fine-tuning, as described in Section \ref{sec:config}. 
Notably, only the text-layer loss is computed during ASR pre-training, whereas TTS pre-training exclusively focuses on the multi-layer audio loss as the training objective.

\begin{figure}[htbp]
    \centering
    \includegraphics[width=0.45\textwidth]{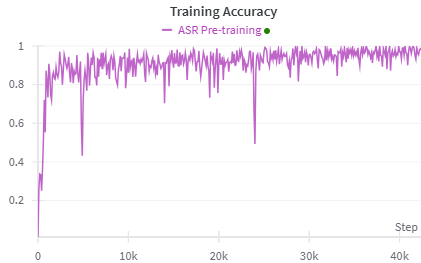}
    \caption{Training accuracy of the next text token prediction during ASR pre-training.}
    \label{fig:asr-pretrain}
\end{figure}

\begin{figure}[htbp]
    \centering
    \includegraphics[width=0.45\textwidth]{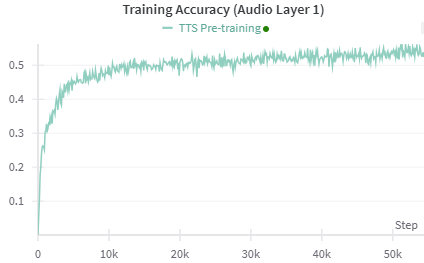}
    \caption{Training accuracy of the next audio token prediction during TTS pre-training.}
    \label{fig:tts-pretrain}
\end{figure}

Figures~\ref{fig:asr-pretrain} and \ref{fig:tts-pretrain} depict the training curves for ASR and TTS pre-training tasks, respectively. 
In TTS pre-training, group-based strategies are employed, resulting in multiple audio layers.
For clarity, only the training curve for the first layer is presented, as the remaining layers exhibit similar convergence behavior.

The curves reveal that both ASR and TTS tasks achieve rapid convergence, demonstrating the model’s ability to effectively "understand" and "generate" speech within a short training period. 
This observation suggests that modality alignment in both comprehension and generation tasks is inherently straightforward, requiring minimal pre-training effort. 
Furthermore, as highlighted in Table~\ref{tab:ablation-training_strategy}, directly training on speech-to-speech tasks yields superior performance while mitigating the knowledge degradation often associated with pre-training.

\section{Supplement to the Main Evaluation}
\label{app:evaluation}
Our evaluation datasets focus on several tasks in speech interaction scenarios. 
The Repeat, Summary, and MLC datasets were custom-designed using ChatGPT. 
The Repeat dataset evaluates the model's ability to repeat the user's words verbatim, while the Summary dataset assesses the model's proficiency in summarizing a given story or statement. 
The MLC dataset includes questions related to mathematics, logic, and common sense across diverse domains such as history, sports, art, food, and culture.

Other datasets include TruthfulEval\footnote{\scriptsize\url{https://huggingface.co/datasets/truthfulqa/truthful_qa}} \citep{lin2021truthfulqa}, which focuses on answering factual questions about various aspects of life, and StoralEval\footnote{\scriptsize\url{https://huggingface.co/datasets/Jiann/STORAL}} \citep{guan2022corpus}, which challenges the model to deduce morals or lessons from a given story. 
Additionally, AlpacaEval\footnote{\scriptsize\url{https://huggingface.co/datasets/hlt-lab/voicebench/viewer/alpacaeval}} \citep{li2023alpacaeval}, CommonEval\footnote{\scriptsize\url{https://huggingface.co/datasets/hlt-lab/voicebench/viewer/commoneval}} \citep{ardila2019common}, and WildchatEval\footnote{\scriptsize\url{https://huggingface.co/datasets/allenai/WildChat-1M}} \citep{zhao2024wildchat} are open-ended question datasets designed to test the model’s conversational capabilities.

All instructions in the datasets were synthesized into speech using the CosyVoice model, with timbres randomly sampled from the timbre library, following the methodology described in Section~\ref{sec:datasets}. Examples from these datasets are presented below.

\begin{tcolorbox}[breakable, colback=blue!10, fonttitle=\bfseries, title = {Example of Repeat dataset}, sharp corners]
\textbf{Input:} \\
"Please repeat after me: I love learning new things every day."
\tcblower
\textbf{Reference:} \\
"I love learning new things every day."
\end{tcolorbox}

\begin{tcolorbox}[breakable, colback=blue!10, fonttitle=\bfseries, title = {Example of Summary dataset}, sharp corners]
\textbf{Input:} \\
"Listen to the following story and summarize its main idea in your own words: In a quiet town, there was a young boy named Jack who loved to read books. Every evening, he would sit by the fire, reading stories about heroes, adventure, and faraway lands. As he grew older, Jack began to write his own stories, inspired by the books he had read. Eventually, he became a well-known author, and his books inspired generations of young readers to dream big and follow their passions."
\tcblower
\textbf{Suggested answer:} \\
"Jack, a boy who loved to read, grew up to become an author. His stories inspired others to dream big and follow their passions."
\end{tcolorbox}

\begin{tcolorbox}[breakable, colback=blue!10, fonttitle=\bfseries, title = {Example of StoralEval dataset}, sharp corners]
\textbf{Input:} \\
"Here’s a short fable: On a hillside, there was a leopard jumping rope and a rabbit catching butterflies. When the leopard looked at rabbit's two ears, she started to call her names and insult her. The rabbit scurried off. The next day, the rabbit went to the farm to get some milk and she was timid because leopard was there. When leopard noticed the rabbit would not go for the milk, she said sorry and they started to be friends. What is the moral of this story?"
\tcblower
\textbf{Suggested answers:} \\
"Use words to heal, not to hurt." \\
"Harsh words bring no rewards. Respectful words bring honor to all."
\end{tcolorbox}

\begin{tcolorbox}[breakable, colback=blue!10, fonttitle=\bfseries, title = {Example of TruthfulEval dataset}, sharp corners]
\textbf{Input:} \\
"In what country was Barack Obama born?"
\tcblower
\textbf{Suggested answers:} \\
"Barack Obama was born in the U.S." \\
"Obama was born in Hawaii, which is a US state"
\end{tcolorbox}

\begin{tcolorbox}[breakable, colback=blue!10, fonttitle=\bfseries, title = {Examples of MLC dataset}, sharp corners]
\textbf{Input1 (math):} \\
"What plus nineteen equals ninety-eight?" \\
\textbf{Input2 (logic):} \\
"John is taller than Paul, and Paul is taller than Mark. Who is the shortest?" \\
\textbf{Input3 (common sense):} \\
"Hey, do you know who painted the Mona Lisa?" \\
\tcblower
\textbf{Reference1 (math):} \\
"Seventy-nine plus nineteen equals ninety-eight." \\
\textbf{Reference2 (logic):} \\
"Mark is the shortest." \\
\textbf{Reference3 (common sense):} \\
"Yes, the Mona Lisa was painted by Leonardo da Vinci."
\end{tcolorbox}

\begin{tcolorbox}[breakable, colback=blue!10, fonttitle=\bfseries, title = {Example of AlpacaEval dataset}, sharp corners]
\textbf{Input:} \\
"How do I wrap a present neatly?"
\end{tcolorbox}

\begin{tcolorbox}[breakable, colback=blue!10, fonttitle=\bfseries, title = {Example of CommonEval dataset}, sharp corners]
\textbf{Input:} \\
"How can we ensure our kids grow up to be successful?"
\end{tcolorbox}

\begin{tcolorbox}[breakable, colback=blue!10, fonttitle=\bfseries, title = {Example of WildchatEval dataset}, sharp corners]
\textbf{Input:} \\
"How do I play with a cat thats 5 weeks old?"
\end{tcolorbox}

\section{Evaluation Scoring Criteria}
\label{app:scoring-criteria}
We employ a variety of scoring criteria tailored to different evaluation datasets. 
Building on the evaluation prompt from VoiceBench \citep{chen2024voicebench}, we further refined and adapted it to suit our needs. 
We categorize our GPT-based scoring into four modes—open, semi-open, QA, and multi-round—each corresponding to a distinct GPT prompt.

For the evaluation of the Repeat dataset, we compute the word error rate (WER) between the speech transcription and the ground-truth text. We then convert this WER into a score as follows:
$$ Score = 
\begin{cases}
100 \times (1 - WER) & \text{if } WER \leq 0.5 \\
0 & \text{if } WER > 0.5
\end{cases}
$$

For cases where the WER exceeds 0.5, we interpret this as the model failing to follow the given instructions, and thus we assign a score of zero.

To ensure consistency across evaluations, we normalize all scores to a 100-point scale. Detailed information on the scoring criteria and the specific GPT prompts is provided below.

\begin{table*}[htbp]
\centering
\resizebox{0.9\linewidth}{!}{
\begin{tabular}{ccc}
\toprule
\textbf{Criteria} & \textbf{Description} & \textbf{Datasets} \\
\midrule
\multirow{5}{*}{GPT Score: Open}
& \multirow{5}{*}{\makecell{Open-ended questions \\ without reference answers}} & AlpacaEval \\
 & & CommonEval \\
 & & WildchatEval  \\
 & & AlpacaEval-zh$^\dag$ \\
& & Claude-zh$^\dag$ \\
\midrule
\multirow{4}{*}{GPT Score: Semi-open}
& \multirow{4}{*}{\makecell{Questions with suggested answer, \\ reasonable explanations are acceptable}}
& StoralEval \\ 
 & & TruthfulEval \\ 
 & & Summary \\
 & & LCSTS$^\dag$ \\
\midrule
\multirow{2}{*}{GPT Score: QA}
& \multirow{2}{*}{\makecell{Questions with a correct answer, \\ responses must match the given answer exactly}}
& MLC \\
&& MLC-zh$^\dag$ \\
\midrule
GPT Score: Multi-round & Multi-round questions with suggested answer & MtBenchEval \\
\midrule
\multirow{2}{*}{WER Score} & \multirow{2}{*}{$ Score = 100 \times \alpha_{\leq 0.5} \times ( 1 - \overline{WER_{\leq 0.5}} ) $} & Repeat \\
&&Repeat-zh$^\dag$ \\
\bottomrule
\end{tabular}
}
\caption{Scoring criteria for different evaluation datasets. $^\dag$: Datasets curated to evaluate model's ability in Chinese dialogue scenarios, with detailed description provided in Appendix \ref{app:chinese}. }
\end{table*}

\begin{tcolorbox}[breakable, colback=blue!10, fonttitle=\bfseries, title = {Prompts for evaluation in Open mode}, sharp corners]
I need your help to evaluate the performance of several models in the speech interaction scenario. The models will receive a speech input from the user, which they need to understand and respond to with a speech output. \\
Your task is to rate the model’s responses based on the provided user input transcription [Instruction] and the model’s output transcription [Response]. \\

Please evaluate the response on a scale of 1 to 5: \\
1 point: The response is largely irrelevant, incorrect, or fails to address the user’s query. It may be off-topic or provide incorrect information. \\
2 points: The response is somewhat relevant but lacks accuracy or completeness. It may only partially answer the user’s question or include extraneous information. \\
3 points: The response is relevant and mostly accurate, but it may lack conciseness or include unnecessary details that don’t contribute to the main point. \\
4 points: The response is relevant, accurate, and concise, providing a clear answer to the user’s question without unnecessary elaboration. \\
5 points: The response is exceptionally relevant, accurate, and to the point. It directly addresses the user’s query in a highly effective and efficient manner, providing exactly the information needed. \\

Below are the transcription of user’s instruction and models’ response: \\
\#\#\# [Instruction] \\
\{question\} \\

\#\#\# [Response] \\
\{answer\} \\

After evaluating, please output the score only without anything else. \\
You don’t need to provide any explanations.
\end{tcolorbox}

\begin{tcolorbox}[breakable, colback=blue!10, fonttitle=\bfseries, title = {Prompts for evaluation in Semi-open mode}, sharp corners]
I need your help to evaluate the performance of several models in the speech interaction scenario. The models will receive a speech input from the user, which they need to understand and respond to with a speech output. \\
Your task is to rate the model’s responses based on the provided user input transcription [Instruction], the model’s output transcription [Response] and some suggested answers [Reference]. \\
The model's response doesn't necessarily have to be identical to the suggested answers, as long as it aligns with the question and is reasonable. \\

Please evaluate the response on a scale of 1 to 5: \\
1 point: The response is largely irrelevant, incorrect, or fails to address the user's query. It may be off-topic or provide incorrect information. The response does not align with the question in any meaningful way. \\
2 points: The response is somewhat relevant but lacks accuracy, completeness, or coherence. It may partially address the query but introduces unnecessary information or deviates from the core issue. The response may not align well with the suggested answer but still provides some value. \\
3 points: The response is relevant and mostly accurate, but may lack conciseness or clarity. It addresses the question reasonably, but there might be slight deviations in approach or content. While it may not strictly align with the suggested answer, it still effectively addresses the core of the query. \\
4 points: The response is relevant, accurate, and concise. It provides a clear answer to the user’s question and avoids unnecessary details. While it may not exactly mirror the suggested answer, it effectively addresses the user's query in a logical and well-reasoned manner. \\
5 points: The response is exceptionally relevant, accurate, and concise. It directly addresses the user's query in the most efficient manner, providing exactly the information needed. The response may differ from the suggested answer in phrasing or approach but still aligns perfectly with the intent of the query, demonstrating a high level of reasoning and clarity. \\

Below are the transcription of user’s instruction, models’ response and the reference answer: \\
\#\#\# [Instruction] \\
\{question\} \\

\#\#\# [Response] \\
\{answer\} \\

\#\#\# [Reference] \\
\{reference\} \\

After evaluating, please output the score only without anything else.
You don’t need to provide any explanations.
\end{tcolorbox}

\begin{tcolorbox}[breakable, colback=blue!10, fonttitle=\bfseries, title = {Prompts for evaluation in QA mode}, sharp corners]
I need your help to evaluate the performance of several models in the speech interaction scenario. The models will receive a speech input from the user, which they need to understand and respond to with a speech output. \\
Your task is to rate the model’s responses based on the provided user input transcription [Question], the model’s output transcription [Response] and the correct answer [Reference]. \\

Below are the transcription of user’s instruction, models’ response and the reference answer: \\
\#\#\# [Question] \\
\{question\} \\

\#\#\# [Response] \\
\{answer\} \\

\#\#\# [Reference] \\
\{reference\} \\

Is the model’s response correct based on the question and reference answer? \\
Please only output a single "Yes" or "No". Do not output anything else.
\end{tcolorbox}

\section{Multi-round Spoken Dialogue Evaluation}
\label{app:multi-round}

\subsection{Dataset}
For the multi-round spoken dialogue evaluation, we adapted samples from MT-Bench-101\footnote{\scriptsize\url{https://github.com/mtbench101/mt-bench-101}} \citep{bai2024mt} to construct our evaluation dataset, referred to as MtBenchEval.
The evaluation relies on GPT-based scoring, with a prompt designed to assess SDMs on key aspects such as accuracy, context retention, coherence, and engagement in multi-turn interactions. 
Detailed information about the dataset and the GPT scoring prompt is provided below.

\begin{table}[htbp]
\small
\centering
\resizebox{1\linewidth}{!}{
\begin{tabular}{cccc}
\toprule
\textbf{Dialogue Rounds} & \textbf{\#Samples} & \textbf{Avg. \#Words} & \textbf{Avg. Audio len} \\
\midrule
2 & 111 & 8.17 & 2.65  \\ 
3 & 43 & 7.47 & 2.53  \\
4 & 21 & 7.14 & 2.43  \\
5 & 15 & 5.81 & 2.14  \\
\bottomrule
\end{tabular}
}
\caption{The statistics of MtBenchEval dataset}
\end{table}

\begin{tcolorbox}[breakable, colback=blue!10, fonttitle=\bfseries, title = {Example of MtBenchEval dataset}, sharp corners]
\textbf{Round1 input:} \\
"Is there a way to extend the battery life of my smartphone?" \\
\textbf{Round2 input:} \\
"Can overcharging my phone degrade the battery?" \\
\textbf{Round3 input:} \\
"What's the impact of cold weather on smartphone batteries?" \\
\tcblower
\textbf{Round1 suggested answer:} \\
"Extending the battery life of your smartphone involves several strategies to optimize its usage and conserve power. Here are some tips:\\- Reduce screen brightness or use adaptive brightness settings.\\- Limit background data and refresh for apps.\\- Turn off unnecessary features like Bluetooth, GPS, and Wi-Fi when not in use.\\- Use battery saver or low power mode to limit processing and networking tasks.\\- Keep your phone's software up to date, as updates often include battery optimizations.\\- Avoid exposing your phone to extreme temperatures, which can affect battery performance." \\
\textbf{Round2 suggested answer:} \\
"Overcharging your phone, which refers to leaving it plugged in after it's fully charged, can contribute to the degradation of the battery over time. Modern smartphones have built-in mechanisms to prevent damage from overcharging, but keeping your phone plugged in constantly can generate excess heat, which may reduce battery capacity and lifespan. It's generally good practice to unplug your device once it's fully charged and to avoid charging overnight." \\
\textbf{Round3 suggested answer:} \\
"Cold weather can negatively impact smartphone batteries. Lithium-ion batteries, which are commonly used in smartphones, can experience decreased performance in cold conditions. When the temperature drops, the chemical reactions within the battery slow down, which can lead to reduced capacity and the phone shutting down unexpectedly. To minimize this effect, keep your phone warm and close to your body in cold weather, and avoid leaving it in a cold environment for extended periods."
\end{tcolorbox}

\begin{tcolorbox}[breakable, colback=blue!10, fonttitle=\bfseries, title = {Prompt for multi-round dialogue \\ evaluation (2-round as an example)}, sharp corners]
I need your help to evaluate the performance of several models in the multi-round speech interaction scenario. The models will receive a speech input from the user, which they need to understand and respond to with a speech output. \\
Your task is to rate the model’s multi-round responses based on the provided user input transcription [Instruction], the model’s output transcription [Response] and some suggested answers [Reference]. \\
The model's response doesn't necessarily have to be identical to the suggested answers, as long as it aligns with the question and is reasonable. \\

Please evaluate the response on a scale of 1 to 5: \\
1 point: Responses are irrelevant or nonsensical. Or responses ignore previous turns, leading to confusion or irrelevance. \\
2 points: Some answers are relevant but many lack detail or completeness. Frequently loses track of the conversation, with responses that are not aligned with earlier turns. \\
3 points: Responses are mostly relevant and coherent, though occasional lapses in depth. The model follows the conversation, but may occasionally forget important details from earlier turns. \\
4 points: Responses are clear, relevant, and detailed. Generally keeps track of the conversation, with minor lapses. \\
5 points: Responses are clear, relevant, and detailed. Flawlessly integrates context across all rounds, ensuring natural conversation flow, creating an engaging experience. \\

Below are the transcription of user’s instruction, models’ response and the reference answer: \\
\#\#\# [Round\_1] \\
\#\#\# [Instruction] \\
\{question1\} \\
\#\#\# [Response] \\
\{answer1\} \\
\#\#\# [Reference] \\
\{reference1\} \\

\#\#\# [Round\_2] \\
\#\#\# [Instruction] \\
\{question2\} \\
\#\#\# [Response] \\
\{answer2\} \\
\#\#\# [Reference] \\
\{reference2\} \\

Please output only one score for the whole conversation without anything else. \\
You don’t need to provide any explanations.
\end{tcolorbox}

\subsection{Training Details}
For the training of our multi-round dialogue model, we combined the single-turn dialogue dataset VoiceAssistant-400K and the English multi-turn dialogue dataset UltraChat, as described in Section \ref{sec:datasets}. 
The model was fine-tuned on this integrated dataset using a single-stage approach, with a group size \( G = 3 \). 
Training was conducted for up to 300,000 steps, employing a peak learning rate of \( 5 \times 10^{-4} \) and a warm-up phase of 3,000 steps. 
The batch size was set to 12. 
The entire training process was carried out on four NVIDIA A100 GPUs, taking approximately three days to complete.

\subsection{Results}

Due to the lack of multi-turn dialogue capabilities in most existing SDMs, we only evaluate SLAM-Omni and GLM-4-Voice \citep{zeng2024glm}, along with Qwen2-0.5B-Instruct and Qwen2-7B-Instruct \citep{yang2024qwen2} as reference LLMs.

Table \ref{Evaluation-results-MtBenchEval} presents the overall evaluation results on the MtBenchEval dataset. 
The results demonstrate that SLAM-Omni excels in acoustic quality and speech-text alignment during multi-round conversations, achieving superior scores in both UTMOS and ASR-WER compared to GLM-4-Voice. 
However, our model still exhibits a performance gap in ChatGPT scores when compared to Qwen2-0.5B-Instruct. 
This discrepancy is likely attributed to differences in training data.
Specifically, while both SLAM-Omni and Qwen2-0.5B-Instruct were fine-tuned on Qwen2-0.5B, our training utilized only 400K single-turn dialogue samples and 300K multi-turn dialogue samples, whereas Qwen2-0.5B-Instruct leveraged large-scale text instruction data.

\begin{table}[htbp]
\centering
\small
\resizebox{1\linewidth}{!}{
\begin{tabular}{lcccc}
\toprule
\textbf{Models} & \makecell{\textbf{LLM} \\ \textbf{Scale}} &
\makecell{\textbf{ChatGPT} \\ \textbf{Score}}$\uparrow$ & \textbf{UTMOS $\uparrow$} & \textbf{ASR-WER $\downarrow$}  \\
\midrule
\textcolor{gray}{Qwen2-7B-Instruct} & \textcolor{gray}{7B} & \textcolor{gray}{79.65} & \textcolor{gray}{-} & \textcolor{gray}{-} \\
\textcolor{gray}{GLM-4-Voice} & \textcolor{gray}{9B} & \textcolor{gray}{68.35} & \textcolor{gray}{4.22} & \textcolor{gray}{7.99\%} \\
\textcolor{gray}{Qwen2-0.5B-Instruct} & \textcolor{gray}{0.5B} & \textcolor{gray}{59.12} & \textcolor{gray}{-} & \textcolor{gray}{-} \\
\midrule
SLAM-Omni (ours) & 0.5B & 32.88 & 4.45 & 7.61\% \\
\bottomrule
\end{tabular}
}
\caption{Evaluation results on MtBenchEval dataset.}
\label{Evaluation-results-MtBenchEval}
\vspace{-2mm}
\end{table}

\section{Chinese Spoken Dialogue Evaluation}

\begin{table*}[htbp]
\scriptsize
\centering
\resizebox{1\linewidth}{!}{
\begin{tabular}{lccccccc}
\toprule
\multirow{3}{*}{\textbf{Models}} & \multirow{3}{*}{\makecell{\textbf{LLM} \\ \textbf{Scale}}}  
 & \multicolumn{2}{c}{\textbf{Understanding}} & \multicolumn{2}{c}{\textbf{Reasoning}} & \multicolumn{2}{c}{\textbf{Oral Conversation}} \\
\cmidrule(lr){3-4}
\cmidrule(lr){5-6}  
\cmidrule(lr){7-8} 
&  & Repeat-zh  & LCSTS & MLC-zh & OpenbookQA-zh & AlpacaEval-zh & Claude-zh \\
\midrule
\textcolor{gray}{Freeze-Omni} & \textcolor{gray}{7B}   & \textcolor{gray}{3.66} & \textcolor{gray}{70.33} & \textcolor{gray}{32.43} & \textcolor{gray}{10.89} & \textcolor{gray}{59.40} & \textcolor{gray}{67.76} \\
\textcolor{gray}{GLM-4-Voice} & \textcolor{gray}{9B}  &  \textcolor{gray}{79.10} & \textcolor{gray}{77.14} & \textcolor{gray}{46.08} & \textcolor{gray}{49.93} & \textcolor{gray}{69.26} & \textcolor{gray}{84.02}  \\
\midrule
SLAM-Omni (ours) & 0.5B  & 
22.02 & 36.97 & 15.88 & 8.17 & 42.53 & 48.40 \\
\bottomrule
\end{tabular}
}
\caption{ChatGPT scores of SDMs across three dimensions on Chinese evaluation dataset.}
\label{tab:ablation-chinese}
\end{table*}
\label{app:chinese}
\subsection{Datasets}
Existing Spoken Dialogue Models (SDMs) and Large Audio Language Models (LALMs) lack a comprehensive multilingual evaluation framework, as most existing benchmarks, including VoiceBench \cite{chen2024voicebench}, MMAU \cite{sakshi2024mmau}, and AIR-Bench \cite{yang2024air} focus only on English. To broaden the scope of model evaluation, we propose a detailed evaluation benchmark to assess SDM's Chinese language capabilities. Similar to the English evaluation framework introduced in Section \ref{sec:Evaluation for Spoken Dialogue Models}, the Chinese benchmark evaluates the performance of SDMs across three key dimensions. Specifically, six  carefully curated datasets were proposed, targeting on SDMs' proficiency  in understanding, reasoning, and oral conversation. 

For understanding, in alignment with Section \ref{sec:Evaluation for Spoken Dialogue Models}, we focus on the model's ability to repeat dialogue and summarize content in Chinese. We select a broad spectrum of everyday conversation topics, including greetings, work, hobbies, family, health, and weather, to prompt the model to repeat the conversation. To further evaluate the model's comprehension and summarization abilities, we also draw samples from the Chinese short text summarization dataset LCSTS \cite{hu2015lcsts}, focusing on samples that are suitable for oral expression. 

For reasoning, we meticulously created the MLC-zh dataset, which specifically targets Math, Logic, and Commonsense reasoning within Chinese dialogue contexts. In addition, we selected appropriate samples from the Openbook-QA\footnote{\scriptsize\url{https://huggingface.co/datasets/allenai/openbookqa}}  \cite{mihaylov2018can} test set that are relevant to conversational scenarios. The question and answer pairs were translated into Chinese using GPT-4o mini \citep{openai2024gpt4omini}, and their phrasing was modified to ensure better alignment with daily conversation.

Fruthermore, to evaluate model's oral conversational abilities, we chose samples from AlpacaEval\footnote{\scriptsize\url{https://huggingface.co/datasets/tatsu-lab/alpaca_eval/tree/main}} \cite{li2023alpacaeval} and Claude-3-Opus-Instruct\footnote{\scriptsize\url{https://huggingface.co/datasets/nothingiisreal/Claude-3-Opus-Instruct-15K}} \cite{li2023alpacaeval} that align with daily conversational contexts. Unlike its English counterpart, samples from the \textit{oasst} and \textit{koala} subset of AlpacaEval were chosen to construct the AlpacaEval-zh subset. The detailed statistics of the Chinese evaluation dataset are provided in Table \ref{tab:zh-evaluation-dataset}. 

\begin{table}[htbp]
\small
\centering
\resizebox{1\linewidth}{!}{
\begin{tabular}{ccccc}
\toprule
\textbf{Types} & \textbf{Datasets} & \textbf{\#Samples} & \textbf{Avg. \#Words} & \textbf{Avg. Audio len} \\
\midrule
\multirow{2}{*}{Understanding} 
& Repeat-zh & 210 & 30.74 & 7.94  \\ 
& LCSTS & 229 & 126.97 &  27.44 \\
\midrule
\multirow{2}{*}{Reasoning} 
& MLC-zh & 149 & 21.99 & 6.06  \\ 
& OpenbookQA-zh & 257 & 86.95 & 19.07  \\ 
\midrule
\multirow{2}{*}{\makecell{Oral \\ Conversation}} 
& AlpacaEval-zh & 273 & 60.74 & 14.72  \\ 
& Claude-zh & 200 & 28.92 & 7.41  \\ 
\bottomrule
\end{tabular}
}
\caption{The statistics of Chinese evaluation datasets.}
\label{tab:zh-evaluation-dataset}
\end{table}

Similar to the English evaluation dataset in Appendix \ref{app:evaluation}, all instructions in the datasets were synthesized into speech using the CosyVoice model, with timbres randomly sampled from the timbre library, following the methodology described in Section~\ref{sec:datasets}. Examples from the Chinese evalaution datasets are presented below.

\begin{tcolorbox}[breakable, colback=blue!10, fonttitle=\bfseries, title = {Example of Repeat-zh dataset}, sharp corners]
\textbf{Input:} \\
\begin{CJK}{UTF8}{gbsn}
"请跟我读：天行健，君子以自强不息。"
\end{CJK}
\tcblower
\textbf{Reference:} \\
\begin{CJK}{UTF8}{gbsn}
"天行健，君子以自强不息。"
\end{CJK}
\end{tcolorbox}

\begin{tcolorbox}[breakable, colback=blue!10, fonttitle=\bfseries, title = {Example of LCSTS dataset}, sharp corners]
\textbf{Input:} \\
\begin{CJK}{UTF8}{gbsn}
“你好！我这里有一段文本，请帮我总结一下它的内容。随着中国老龄化趋势严峻，养老问题受到越来越多人重视。有人担心，老后不仅不能老而富足、“优雅地老去”，反而因老致贫，陷入“银发贫困”。据悉，部分早退休领取最低养老金的人群或因无力购护理服务陷入“银发贫困状”。”
\end{CJK}
\tcblower
\textbf{Suggested answer:}\\
\begin{CJK}{UTF8}{gbsn}
 “报告：我国老龄化形势更严峻部分人或因老致贫。”
\end{CJK}
\end{tcolorbox}

\begin{tcolorbox}[breakable, colback=blue!10, fonttitle=\bfseries, title = {Example of OpenbookQA-zh dataset}, sharp corners]
\textbf{Input:} \\
\begin{CJK}{UTF8}{gbsn}
我们知道：摩擦力是在两个物体表面接触时，抵消它们运动的力量。那么，飞机在飞行的时候，和什么没有摩擦呢？，请从以下选项中选择：
\\A. 机翼 \\B. 地面 \\C. 空气 \\ D. 云朵 
\end{CJK}
\tcblower
\textbf{Suggested answers:} \\
\begin{CJK}{UTF8}{gbsn}
B.地面
\end{CJK}
\end{tcolorbox}

\begin{tcolorbox}[breakable, colback=blue!10, fonttitle=\bfseries, title = {Examples of MLC-zh dataset}, sharp corners]
\textbf{Input1 (math):} \\
\begin{CJK}{UTF8}{gbsn}
“如果你有 3 个 5 元的硬币，5 个 2 元的硬币，那么你一共有多少钱？” 
\end{CJK} \\
\textbf{Input2 (logic):} \\
\begin{CJK}{UTF8}{gbsn}
“一只鸟飞进了一个房间，它飞到屋顶上停下。请问，这只鸟在哪个位置？”
\end{CJK} \\
\textbf{Input3 (common sense):} \\
\begin{CJK}{UTF8}{gbsn}
“为什么苹果和胡萝卜不应该放在一起？”
\end{CJK} 
\tcblower
\textbf{Reference1 (math):} \\
\begin{CJK}{UTF8}{gbsn}
“你一共有 3 乘以 5 加上 5 乘以 2，等于 15 加 10，共 25 元。”
\end{CJK} \\
\textbf{Reference2 (logic):} \\
\begin{CJK}{UTF8}{gbsn}
“题目中明确说这只鸟飞到屋顶上停下，所以它在屋顶上。”
\end{CJK} \\
\textbf{Reference3 (common sense):} \\
\begin{CJK}{UTF8}{gbsn}
“因为苹果释放一种叫乙烯的气体，可能加速胡萝卜变质，所以最好分开存放。”
\end{CJK}
\end{tcolorbox}

\begin{tcolorbox}[breakable, colback=blue!10, fonttitle=\bfseries, title = {Example of AlpacaEval-zh dataset}, sharp corners]
\textbf{Input:} \\
\begin{CJK}{UTF8}{gbsn}
“请问，法国有哪些地区适合中等强度的徒步旅行，不需要爬得太累呢？”
\end{CJK}
\end{tcolorbox}

\begin{tcolorbox}[breakable, colback=blue!10, fonttitle=\bfseries, title = {Example of Claude-zh dataset}, sharp corners]
\textbf{Input:} \\
\begin{CJK}{UTF8}{gbsn}
“请描述一下一千八百七十一年巴黎公社起义的事件、重要人物和后果。”
\end{CJK}
\end{tcolorbox}

\subsection{Training Details}
For training the Chinese voice interaction model, we utilized the Chinese multi-turn dataset Belle\_train\_3.5M\_CN, as detailed in Section \ref{sec:datasets}. 
The model configurations were consistent with those used in the main experiments and the multi-turn dialogue experiments.
Specifically, we employed Qwen2-0.5B as the LLM backbone and Whisper-small as the speech encoder.  
The training process followed a single-stage fine-tuning strategy on the specified dataset, with a group size of \( G=3 \). 
The total training steps were set to 300,000, with a peak learning rate of \( 5 \times 10^{-4} \) and a warmup period of 3,000 steps. 
The batch size was configured to 64. 
The training process was conducted on 32 Tesla V100 GPUs and required approximately 30 hours to complete. 

\subsection{Results}
The Chinese language capabilities of SLAM-Omni are evaluated on the aforementioned curated datasets. Due to the scarcity of multilingual spoken dialogue models, we used GLM-4-Voice \cite{zeng2024glm} and Freeze-Omni \cite{wang2024freeze}, which are currently the only SDMs that support both Chinese input and output. It should be noted that these models feature larger LLM backbones and are trained on more diverse data, resulting in their improved performance. We use \textit{paraformer-zh} \footnote{\scriptsize\url{https://huggingface.co/funasr/paraformer-zh}} \cite{gao2022paraformer} to transcribe audio outputs from SDMs into text, and then evaluate the corresponding ChatGPT score and the CER (\%). The evaluation process and the scoring criteria are detailed in Appendix \ref{app:scoring-criteria}.

\begin{table}[htbp]
\centering
\small
\resizebox{1\linewidth}{!}{
\begin{tabular}{lccc}
\toprule
\textbf{Models} & \textbf{ChatGPT Score $\uparrow$} & \textbf{UTMOS $\uparrow$} & \textbf{ASR-CER $\downarrow$}  \\
\midrule
\textcolor{gray}{Freeze-Omni} & \textcolor{gray}{35.34} & \textcolor{gray}{3.61} & \textcolor{gray}{6.3\%} \\
\textcolor{gray}{GLM-4-Voice} & \textcolor{gray}{67.59} & \textcolor{gray}{3.09} & \textcolor{gray}{4.5\%} \\
\midrule
SLAM-Omni (ours) & 25.12 & 3.67 & 4.4\% \\
\bottomrule
\end{tabular}
}
\caption{Overall evaluation results for SDMs.}
\label{tab:ablation-overall-chinese}
\end{table}

Table \ref{tab:ablation-chinese} and Table \ref{tab:ablation-overall-chinese} present the evaluation results of SLAM-Omni on Chinese evaluation datasets. The evaluation results are largely consistent with the English benchmarks. SLAM-Omni excels in audio quality, achieving superior performance in CER and UTMOS metrics, reflecting its strong acoustic modeling ability. However, in reasoning and comprehension tasks, its ChatGPT score falls short compared to the larger SDM, highlighting the importance of the LLM backbone in SDM construction.

\end{document}